\documentclass[11pt,epsf,letterpaper]{article}%
\usepackage{setspace}
\usepackage{color}
\usepackage{graphicx}
\usepackage{amsmath}
\usepackage{amsfonts}
\usepackage{verbatim}
\usepackage{amssymb}

\usepackage{url}

\providecommand{\U}[1]{\protect\rule{.1in}{.1in}}
\newcommand{\be}{\begin{equation}}
\newcommand{\ee}{\end{equation}}
\newcommand{\ba}{\begin{eqnarray}}
\newcommand{\ea}{\end{eqnarray}}

\begin{document}

\title{\bf Searching for a response: the intriguing mystery of Feynman's theoretical reference amplifier}

\author{Vincenzo d'Alessandro$^{1}$,
Santolo Daliento$^{1}, $Marco Di Mauro$^{2,3}$,
Salvatore Esposito$^{3}$, \\
Adele Naddeo$^{3}$ \\
$^{1}$\footnotesize{Dipartimento di Ingegneria Elettrica e delle Tecnologie dell'Informazione, Universit\'a di Napoli ``Federico II",} \\ \footnotesize{Via Claudio 21, 80125 Naples, Italy.}\\
$^{2}$\footnotesize{Dipartimento di Fisica ``E.R. Caianiello",
Universit\'a di Salerno,
Via Giovanni Paolo II, 84084 Fisciano, Italy.}\\
$^{3}$\footnotesize{INFN Sezione di Napoli, Via Cinthia, 80126 Naples, Italy.}}
\maketitle

\begin{abstract}
\noindent We analyze Feynman's work on the response of an amplifier performed at Los Alamos and described in a technical report of 1946, as well as lectured on at the Cornell University in 1946-47 during his course on Mathematical Methods. The motivation for such a work was Feynman's involvement in the Manhattan Project, for which the necessity emerged of feeding the output pulses of counters into amplifiers or several other circuits, with the risk of introducing distortion at each step. In order to deal with such a problem, Feynman designed a theoretical ``reference amplifier", thus enabling a characterization of the distortion by means of a benchmark relationship between phase and amplification for each frequency, and providing a standard tool for comparing the operation of real devices. A general theory was elaborated, from which he was able to deduce the basic features of an amplifier just from its response to a pulse or to a sine wave of definite frequency. Moreover, in order to apply such a theory to practical problems, a couple of remarkable examples were worked out, both for high-frequency cutoff amplifiers and for low-frequency ones. A special consideration deserves a mysteriously exceptional amplifier with best stability behavior introduced by Feynman, for which different physical interpretations are here envisaged. Feynman's earlier work then later flowed in the Hughes lectures on Mathematical Methods in Physics and Engineering of 1970-71, where he also remarked on causality properties of an amplifier, that is on certain relations between frequency and phase shift that a real amplifier has to satisfy in order not to allow output signals to appear before input ones. Quite interestingly, dispersion relations to be satisfied by the response function were introduced.
\end{abstract}

\section{Introduction}

Starting from March 1943, shortly after graduating from Princeton,
Richard P. Feynman was involved in the Manhattan project
\cite{Galison,Hoddeson,FeynmanLABelow} for studying a number of
different problems directly related or not to the making of the
bomb.\footnote{{In particular Ref. \cite{Galison} gives a
detailed list of the projects Feynman was involved in.}} To
begin with, he was initially involved in studying instruments and
experimental devices, such as -- for instance -- an experiment
conducted by Thoma Snyder on the counting of neutrons emerging
from the bombardment of  $^{235}$U with slow neutrons
\cite{Snyder}, or even the study of the ``water boiler", a small
nuclear reactor designed to experiment on fundamental properties
of the chain reaction \cite{deHoffmann,
FeynmanNeutronChains,deHoffman2}. He was then sent by Oppenheimer
to Oak Ridge as safety supervisor, whose task was basically to
prevent nuclear disasters caused by inexpert technicians managing
a quantity of uranium sufficiently large to reach the critical
mass for starting a chain reaction
\cite{FeynmanOakRidge,FeynmanOakRidge2}. Feynman also developed an
integral theorem that allowed to evaluate the distribution of
neutrons and active material from known distributions, in order to
maximize the number of neutrons leading to a successful chain
reaction \cite{FeynmanIntegralTh}. Furthermore, he had to deal
with numerical calculations concerning implosion plutonium bombs
\cite{FeynmanCompImplosion}, rather than uranium ones, this last
project being assigned to him by the theory division leader Hans
Bethe, whom he would follow to Cornell University after the end of
the war. However, the most important and difficult project (though
never effectively realized) concerned the ``hydride bomb", which
was supposed to work around a uranium hydride (rather than pure
uranium) core, where the hydrogen atom in the hydride would favor
the slowing down of neutrons originating the chain reaction, thus
consuming less $^{235}$U than the ordinary metal bomb
\cite{FeynmanHydride1,FeynmanHydride2,FeynmanHydride3,
FeynmanHydride4}. The difficulty of such study came from the fact
that he had to manage neutron distributions with {\it different}
velocities, by employing more refined mathematical tools
concerning Boltzmann kinetics. As {pointed out by Galison} \cite{Galison}, in all these works -- differently from other
scholars -- Feynman directly focused on the solutions of the
relevant equations, rather than on the equations themselves, an
approach that will be adopted later even in his most famous
contributions in Quantum Electrodynamics.

As a matter of fact, in his work at Los Alamos, Feynman was mainly
involved in technical and engineering issues. In most experiments,
the basic aim was simply to count neutrons emerging from a given
reaction, in order to estimate the efficiency of such reaction,
but the neutron signals were usually so small that an amplifier
was required to study it. The practical problem with feeding the
output pulses of counters into amplifiers and various other
circuits was mainly the emergence of distortions, since, as
Feynman noted {(see Ref. \cite{FeynmanLARep} on page 3)},
amplifiers usually distort signals at very high and very low
frequencies, while they work correctly in between. In order to
solve such a problem, instead of studying the details of the
different amplifiers employed in the different experiments,
Feynman designed a theoretical ``reference amplifier" distorting
the signal either at the low or at the high end of its responsive
range, thus providing a standard to be compared with the real
devices. Indeed, he succeeded in characterizing the distortion
introduced by means of a benchmark relationship between phase and
amplification of the signal for each frequency component. This
interesting work is described in one of his technical reports of
1946 \cite{FeynmanLARep}, and later addressed in lectures given at
the Cornell University in 1946-47. Here, his very first course was
on Mathematical Methods in Physics, a rather standard course
dealing with different mathematical tools \cite{Cornell}, although
a number of curious tricks and interesting points were present. In
particular, a substantial part of this course was devoted to
integration methods and applications of residue theorem, among
which the problem of deriving the response of an amplifier was
considered.\footnote{{Cfr. Ref. \cite{Cornell}, part 1, on pages 167-172.}} 

{In addition to the amplifier, in these lectures
Feynman often drew inspiration from problems he worked out during the
Manhattan project, although sometimes, as in the case of the amplifier, he
apparently did not explicitly quote his source (or at least it is not
specified in the notes). Another interesting issue mentioned in the lectures is
the continued fraction representation of Bessel functions,
\footnote{{Cfr. page 131 of part I of Ref. \cite{Cornell}.}} which was of some
relevance in numerical calculations. Later on, in
delivering the Hughes lectures on Mathematical Methods
(see \cite{Hughes}, on pages 70-71), when talking again about this topic, Feynman
explicitly remarked that  ``it held up the development of the
atomic bomb while we worked it out''. In other places, 
references to neutron absorbers are explicit. For example,
in part I (pages 96-98) of Ref. \cite{Cornell},  a neutron absorber is
studied, while in part II (page 62) a differential equation concerning the
behavior of neutron density is described, the same equation being applied (on page 126) to a ``gadget'', which closely reminds the core of an atomic bomb. All these examples confirm
and complement the thesis put forward by Galison
\cite{Galison} that Feynman's experience in the Manhattan project
strongly shaped his approach to physics, since they clearly
show that also his teaching, as well as his research, was
influenced.} Finally, the same amplifier problem was taken on
later in the lectures he delivered (after he moved to Caltech) at
the Hughes Aircraft Company \cite{noi}, in the 1970-71 course on
Mathematical Methods in Physics and Engineering {(\cite{Hughes} on pages 118-128)}, grossly
following his previous course at Cornell quoted above, though he
specially dwelled on specific points. In the latter lectures the
focus was on causality properties of the transfer function,
succeeding even in deriving the Kramers-Kronig dispersion
relations, whose standard framework (also considered by Feynman)
is the application to the light refractive index, {but
which hold for any linear causal system \cite{Toll:1956cya}}.
{In this discussion, indeed, Feynman cited (on page 126) the
(first edition of the) optics textbook \cite{Lipson}, and then
briefly mentioned possible applications to optics (on page 128).}

In the present paper we {build upon the work in Ref.
\cite{Galison}}, reporting on a complete (technically-assisted)
historical study performed on Feynman's theoretical reference
amplifier, as inferred from what he wrote about it -- directly or
indirectly -- {from} the Manhattan Project time till his
1970-71 Hughes lectures \cite{Hughes}, {passing through the
1946 Cornell lectures \cite{Cornell}}. The general theory
developed by Feynman is highlighted in the following section,
while in Sect. 3, we will dwell shortly on some theoretical issues
he addressed later in his analysis, concerning the causality
properties of the amplifier, as embedded also in dispersion
relations techniques. In Sect. 4, instead, we will analyze a
couple of specific examples considered by Feynman; particularly
interesting will be the discussion of the physical meaning of
those examples, which reveals quite an interesting technical
mystery. Finally, in the concluding section, we will summarize and
discuss what intriguingly emerged in our present study.

\section{Response function of an amplifier}

As said above, during the development of the Manhattan project it was often necessary to amplify signals coming from neutron counters or ionization chambers. Usually, such signals are composed of different frequencies and, when entering an amplifier, amplification is expectedly not the same for all frequency components, thus introducing some distortion in the output signal upon which insight should be gained. Moreover, phase shifts may as well develop for different frequency components, whose behavior as a function of the frequency can be assumed -- as Feynman did -- to be linear to a first approximation \cite{FeynmanLARep}: linearity allowed him to ``sum'' a high pass and a low pass filter in order to have a theoretical ``amplifier" (in the sense specified below) with a behavior similar to that of a real device. Also, the time delay produced in the amplifier can be neglected, so that one can focus just on  distortion.

More in detail \cite{Hughes}, Feynman regarded the amplifier as a black box characterized by the fact that the output voltage $E_\mathrm{OUT}$ is related to the input voltage $E_\mathrm{IN}$ by some quantity $g$ termed the {\it gain} of the device, i.e.
\begin{eqnarray}\label{gain}
E_\mathrm{OUT}=g \, E_\mathrm{IN}.
\end{eqnarray}
Given the assumption of a \textit{linear} amplifier, $g$ is a linear function: if $f_1(t)$ and $f_2(t)$ are two input signals and $F_1(t)$ and $F_2(t)$ are the corresponding output ones, then the sum $f_1 + f_2$ in input gives the output signal $F_1 + F_2$. The amplifier was also assumed to be time-invariant: if at time $t$ the output signal $F(t)$ is obtained from the input one $f(t)$, this same sample signal in input at a later time $t+a$ will produce the same output. A good amplifier is \textit{flat} over a large region of frequencies, that is amplification is nearly independent of frequency in this region, while, on the other hand, for very high and very low frequencies the amplification falls off rapidly. In particular, for high frequencies the amplification follows an inverse power law $\left( {\omega_0}/{\omega} \right)^k$, where $\omega_0$ is some characteristic frequency. Similarly, amplifiers with a low-frequency cutoff have amplification falling off as $\left( {\omega}/{\omega_0} \right)^k$. The high-frequency response affects the shape of a pulse, its rate of rise and the accuracy with which the pulse is followed, the low-frequency counterpart determines instead the response over long times. Feynman performed a different analysis in these two different situations, by considering two kinds of amplifiers: a first one having only a high-frequency cutoff while it is flat for low frequencies, and, conversely, a second one with a low-energy cutoff while passing with unit amplification all frequencies (including very large ones). Due to linearity, the effect of a real amplifier with both cutoffs can be obtained by letting the pulse to pass first through a high-frequency cutoff amplifier and, then, through a second amplifier with a low-frequency cutoff only.

Feynman's peculiar approach was just to consider a very short
signal to be sent  to the amplifier or, more specifically, to
study its response to a delta-function signal, and then
constructing the response to a variety of differently shaped input
signals by considering them as the superposition of a bunch of
delta-functions, each at a given different time and weighted with
a different amplitude. The response function $R(t-t_0)$
corresponding to an input signal that is an  infinitely sharp and
high pulse $\delta(t-t_0)$ is the basic building block of the
theory, satisfying the following correspondence relations between
in and out functions, {again due to linearity}:
\begin{equation}\label{crelations1}
\begin{array}{rcl}
\displaystyle \delta(t-t_0) & \rightarrow & \displaystyle R(t-t_0) , \\
\displaystyle b \, \delta(t-t_0) & \rightarrow & \displaystyle b \, R(t-t_0) , \\
\displaystyle \delta(t-t_1) + \delta(t-t_2) & \rightarrow & \displaystyle  R(t-t_1) + R(t-t_2) , \\
\displaystyle f(t_1) \, \delta(t-t_1) & \rightarrow & \displaystyle F(t_1)  \, R(t-t_1) .
\end{array}
\end{equation}
For a pulse of general shape $f(t)$, written as the superposition
of a {continuous infinity} of delta pulses occurring at
different times, $ f(t)=\int_{-\infty }^{+\infty }f(t^{\prime}) \,
\delta(t-t^{\prime})  \, \mathrm{d}t^{\prime}$, the response of
the amplifier is given by
\begin{eqnarray}\label{goutput}
O(t)=\int_{-\infty }^{+\infty }f(t^{\prime}) \, R(t-t^{\prime}) \, \mathrm{d}t^{\prime},
\end{eqnarray}
$R(t)$ being the response to the single $\delta(t)$ pulse, i.e. a Green's function.

{By considering a sine wave with constant frequency as an input function, namely $E_\mathrm{IN}=\mathrm{e}^{i\omega t}$, then the output one will be a sine wave} with the same frequency,
but amplified and phase shifted, i.e. $E_\mathrm{OUT}=A(\omega) \,
\mathrm{e}^{i\omega t}$ where $A(\omega)=\left| A(\omega)
\right|e^{i\phi(\omega)}$ is the transfer function of the
amplifier {(which is nothing but the Fourier transform of
the response ${R(t-t')}$ to the delta
pulse}\footnote{{In fact, if ${f(t)=\mathrm{e}^{i\omega t}}$, then its Fourier transform is
${\varphi(\omega)=\delta(\omega-\omega ')}$. Therefore
${O(\omega)=\varphi(\omega)A(\omega)=\delta(\omega-\omega
')A(\omega)}$ and
${O(t)=\int_{-\infty}^{\infty}\,\mathrm{d}\,\omega ' \mathrm{e}^{i\omega'
t}\delta(\omega-\omega')A(\omega')=\mathrm{e}^{i\omega t}A(\omega).}$}}):
its magnitude $\left| A(\omega) \right|$ gives the amplification
factor, while the imaginary part yields a phase shift. Feynman's
main focus was just on the quantity $A(\omega)$.

In the general case, an input signal is built of many frequencies,
i.e. $E_\mathrm{IN}=f(t)=\int_{-\infty }^{+\infty } \varphi
\left(\omega \right) \mathrm{e}^{i\omega t} \mathrm{d}\omega$, and
the output will depend on the amplitude of each component, so that
integration over all frequencies is required in order to get the
total output signal:
\begin{eqnarray}\label{generalcase1}
E_\mathrm{OUT}= {O(t)}=\int_{-\infty }^{+\infty } \varphi
\left(\omega \right) A\left(\omega \right) \mathrm{e}^{i\omega t}
\mathrm{d}\omega.
\end{eqnarray}
By introducing the Fourier transform $\varphi \left(\omega \right)= \frac{1 }{2 \pi} \int_{-\infty }^{+\infty }  f(t)  \, \mathrm{e}^{-i\omega t} \mathrm{d}t$, the above expression can be written as:
\begin{eqnarray}\label{generalcase2}
E_\mathrm{OUT}= \int_{-\infty }^{+\infty } \left[ \int_{-\infty
}^{+\infty }  f(t^{\prime})  \, \mathrm{e}^{-i\omega t^{\prime}}
\mathrm{d}t^{\prime} \right] A\left(\omega \right)
\mathrm{e}^{i\omega t} \, \frac{\mathrm{d}\omega }{2 \pi} =
\int_{-\infty }^{+\infty }  f(t^{\prime}) R\left(t-
t^{\prime}\right) \mathrm{d}t^{\prime},
\end{eqnarray}
where, {by comparing with (\ref{goutput}),} we recognize
the Green's function $R(t-t^{\prime})$:
\begin{eqnarray}\label{tf2}
R\left(t-t^{\prime} \right)= \int_{-\infty }^{+\infty } \mathrm{e}^{i\omega (t-t^{\prime})}A(\omega) \, \frac{\mathrm{d}\omega}{2 \pi} \, .
\end{eqnarray}
Eq. (\ref{generalcase2}) involves the convolution of two
functions, $f(t)$ and $R\left(t \right)$, so that it is much more
convenient to switch to the corresponding Fourier transform.
Indeed, as is well known, given two functions $f(t)$ and $g(t)$,
with Fourier transforms $F(\omega)$ and $G(\omega)$ respectively,
in the $\omega$-space their convolution reduces just to
multiplication: $\int  f(t^{\prime}) \, g\left(t-
t^{\prime}\right) \mathrm{d}t^{\prime} \ \rightarrow
F\left(\omega\right) G\left(\omega\right)$. The delta-function
$\delta(t)$ introduced above has a simple Fourier transform,
$\varphi \left(\omega \right)= {1}/{2 \pi}$, whose meaning is that
all frequencies are equally represented and no relative phase
shift is present. By substituting into Eq. (\ref{generalcase1}),
the response function $R(t)$ is then recovered (see Eq.
(\ref{tf2})).

Summing up, Feynman deduced the features of an amplifier from its
response to a pulse or to a sine wave of definite frequency. Given
the general expression for $R(t)$, Feynman's analysis focused on
the behavior of such a function for various choices of
$A\left(\omega \right)$ \cite{FeynmanLARep}. He also briefly
pointed out that a reliable $A\left(\omega \right)$ for a real
amplifier has to satisfy given relations between frequency and
phase shift in order not to allow output signals occurring {\it
before} the introduction of an input signal, i.e. all
singularities (poles and branch points) of $A\left(\omega \right)$
lie on the positive imaginary half of the complex plane. Such a
mathematically-inspired method was inherited from the famous
textbook of 1945 by H.W. Bode \cite{Bode} (which Feynman certainly
knew and used \cite{FeynmanLARep}), originally written as a
technical report for engineers, and subsequently turned into a
book. Later, however, Feynman developed in more detail this issue
in his Hughes lectures on Mathematical Methods in Physics and
Engineering \cite{Hughes}, as we will see in the following
section.

\section{Causality and dispersion relations}

A particularly interesting issue addressed
 by Feynman
\cite{Hughes} was the \textit{causality} properties of his
theoretical reference amplifier, namely the requirement that its
response function $R(\tau)=0$ for $\tau < 0$. Such an issue was
pivotal in Feynman's approach to the amplifier, as apparent from
the fact that it is mentioned in the Los Alamos
report;\footnote{{``There are certain relations between
phase shift and frequency''; cfr. \cite{FeynmanLARep} on page 6.}}
also, from a letter John A. Wheeler wrote to Feynman in 1949, we
see that he considered causality important even before, when just
a PhD student.\footnote{Quoting J.A. Wheeler: ``I remember you
gave a report at Journal Club one Monday evening in 1941 on the
relation between phase change and amplitude gain for a linear
amplifier [...]. The magician was able to deduce all he needed
from the requirement that energy shouldn't come out of the box on
the right hand side before it had been put in on the left"
\cite{WheelerLetter}.}

In general, \textit{strict causality} refers to the fact that no
output can occur before the input, and can be conveniently
expressed in different forms for different physical systems. For a
homogeneous refractive medium, for instance, it can be read as {no
signal can be transmitted faster than the speed of light $c$}.
Causality reflects itself in {\it dispersion relations}, which are
integral formulas relating a dispersive to an absorptive process:
they are ubiquitous in physics, ranging from the theory of light
dispersion in a dielectric medium to the scattering of nuclear
particles \cite{Nuss}, as well as the electrical network theory
\cite{Bode}. A dispersion relation is expected to hold in any
theory where the output function of time is a linear functional of
an input function, the interaction being time-independent, and
where the output function cannot manifest before the application
of the input one.\footnote{The idea of a dispersion relation dates
back to the work of Sommerfeld \cite{Somm} and Brillouin
\cite{Brill}, who showed that, in an idealized dielectric, no
signal travels faster than $c$, although phase velocity and group
velocity may exceed $c$ for some frequencies. By means of analytic
continuation of the refractive index in the complex frequency
plane, Kramers proved that a signal cannot travel faster than $c$
in any medium for which a given dispersion relation is satisfied
\cite{Kramers}. The first proof of the equivalence between
causality and dispersion relations is instead due to Kronig
\cite{Kronig}, though later investigated in detail for
non-relativistic particles \cite{Tiomno,Kampen} as well as for
light \cite{Kampen2}. {The logical equivalence between
strict causality and the validity of dispersion relations in any
linear system was proved in Ref. \cite{Toll:1956cya}.}}

{The importance of dispersion relations in connection with
causality was probably suggested to Feynman again by Bode. It is
well known, in fact, (see e.g. Ref. \cite{bechhoefer}) that Bode
investigated a gain phase version of the Kramers-Kronig relations
in 1937 \cite{Bode 2}, which was also described in his textbook
(it was cited by Feynman in this connection). Despite
this, in the 1946 report and in the Cornell lectures, only a brief
mention to this topic was given. It is not known why Feynman decided to
treat the subject in much more detail in the later Hughes lectures. However, we may speculate
that it is due\footnote{{We thank an anonymous referee for
pointing out to us this possibility.}} to the fact that, during the 1950s and 1960s
(just in the time elapsed between the two sets of lectures), dispersion relations techniques gained
prominence in elementary particle physics, especially S-matrix theory, starting with the work
of Gell-Mann, Goldberger and Thirring \cite{GellMann:1954db} (see
e.g. \cite{Pickering:1985gf,Cushing:1990dt,Wigner} and references
therein). Of course, this did not go unnoticed by Feynman and, indeed, in his Caltech Lectures on General Physics, given in 1961 (see Ref. \cite{FeynmanBook1}, section 31-3), he notably remarks: ``In the past few years, `dispersion equations' have been finding a new use
in the theory of elementary particles''. However, a detailed
answer to this question would require further study, which is outside
the scope of this work.}

Turning back to Feynman's treatment of amplifiers, the requirement
that no response occurs until an input signal is applied
{(see Ref. \cite{Hughes} on pages 125-128)}, translates into the
following equality:
\begin{eqnarray}\label{causality1}
\int_{-\infty }^{0 }R(\tau) \, \mathrm{e}^{-i\omega^{\prime}\tau} \mathrm{d}\tau=0 ,
\end{eqnarray}
and, by substituting Eq. (\ref{tf2}) into the above expression, we have:
\begin{eqnarray}\label{causality2}
\int_{-\infty }^{+\infty } \! \! \int_{-\infty }^{0 }\!\! A(\omega) \, {\mathrm e}^{i(\omega -\omega^{\prime} )\tau} \,  \frac{{\mathrm d} \tau \, {\mathrm d}\omega}{2 \pi}=0.
\end{eqnarray}
The integral over $\tau$ was evaluated by introducing a converging
factor ${\mathrm e}^{\epsilon\tau}$, with $\tau<0$, {(where
it is understood that in the end the limit $\epsilon \rightarrow
0$ will be taken)}, obtaining:
\begin{eqnarray}\label{causality3}
\int_{-\infty }^{0 } e^{i\left(\omega-\omega^{\prime} \right)\tau} e^{\epsilon\tau} d\tau = \lim_{\epsilon \rightarrow 0 } \frac{1}{i\left(\omega-\omega^{\prime} \right)+\epsilon} \, .
\end{eqnarray}
Eq. (\ref{causality2}) then becomes {(multiplying by an
overall ${i}$ factor)}:
\begin{eqnarray}\label{causality4}
\int_{-\infty }^{+\infty } \frac{A(\omega^{\prime})}{\omega^{\prime}-\omega -i\epsilon} \frac{{\mathrm d} \omega^{\prime}}{2 \pi}=0 \, ,
\end{eqnarray}
i.e. a convolution between $A(\omega^{\prime})$ and
{${1}/{(\omega^{\prime}-\omega) }$}, whose Fourier transform with
respect to time is:
\begin{eqnarray}\label{causality5}
\theta \left(-t \right) R\left(t \right)=0
\end{eqnarray}
($\theta \left(-t \right)$ is the unit step function, which is zero for $t<0$). The causality condition can thus be translated by requiring that $A(\omega)$ has no singularities (i.e. poles) below the real axis in the plane of complex frequencies. Now, a given function exhibits a pole for a given complex frequency $\omega= \omega_R + i \omega_I$ when a resonance is present: by approaching the resonant frequency, the oscillation amplitude becomes infinite for a driving force with finite amplitude. Then, according to Feynman \cite{Hughes}, the causality principle suggests that the only way a physical system can achieve an infinite amplitude is as a result of its memory of an infinite driving force at some earlier time. As such, a pole is thus due to a driving force with an exponentially decaying amplitude from time $t=-\infty$ and, as a consequence, the driving force has a complex resonance frequency with positive $\omega_I$, implying that the poles of a real system must lie in the upper half of the complex frequency plane.

Finally, when dealing with the properties of the transfer function $A(\omega)$, Feynman introduced the concept of \textit{dispersion relations} in his discussion. The factor $(\omega^{\prime}-\omega-i\epsilon )^{-1}$ in Eq. (\ref{causality4}) can be conveniently rewritten in terms of its real and imaginary parts as follows:
\begin{eqnarray}\label{sum1}
\frac{1}{\omega^{\prime}-\omega-i\epsilon }=\frac{\omega^{\prime}-\omega}{(\omega^{\prime}-\omega)^2+\epsilon^2 }+\frac{i\epsilon}{(\omega^{\prime}-\omega)^2+\epsilon^2 } \, ,
\end{eqnarray}
so that it is easy to understand its limiting behavior for $\epsilon \rightarrow 0$:
\begin{eqnarray}\label{sum2}
\frac{1}{\omega^{\prime}-\omega-i\epsilon } \rightarrow \mathrm{p.v.} \!\left(\frac{1}{\omega^{\prime}-\omega }\right) + i\pi \delta \left(\omega^{\prime}-\omega \right).
\end{eqnarray}
The causality condition (\ref{causality4}) then becomes:
\begin{eqnarray}\label{sum3}
\int_{-\infty }^{+\infty } \!\! A(\omega^{\prime}) \, \mathrm{p.v.} \!\left( \frac{1}{\omega^{\prime}-\omega } \right) \frac{{\mathrm d}\omega^{\prime}}{ \pi}= -i A(\omega) \, ,
\end{eqnarray}
from which the dispersion relations easily follow for the complex function $A(\omega)=A_R(\omega)+ i A_I(\omega)$:
\begin{eqnarray}\label{dispersion1}
\int_{-\infty }^{+\infty } \!\! A_R(\omega^{\prime}) \, \mathrm{p.v.} \! \left( \frac{1}{\omega^{\prime}-\omega }\right)  \frac{\mathrm d\omega^{\prime}}{ \pi}&=& A_I(\omega),
\\ \nonumber \\
\label{dispersion2}
-\int_{-\infty }^{+\infty } \!\! A_I(\omega^{\prime})  \, \mathrm{p.v.} \! \left( \frac{1}{\omega^{\prime}-\omega }\right) \frac{\mathrm d\omega^{\prime}}{ \pi}&=& A_R(\omega).
\end{eqnarray}
In optics, as Feynman noted, the function $A\left(\omega\right)$
represents the complex refractive index of light: its imaginary
part describes light absorption by a medium, while the real part
gives the frequency-dependent refractive index $n$ (a phenomenon
known as chromatic aberration).
\newpage 

\section{Remarkable examples with a peculiar mismatch}

For high-frequency cutoff amplifiers, Feynman considered a couple
of specific examples, starting from the following transfer
function \cite{Bode},\footnote{{In his report
\cite{FeynmanLARep}, on page 6, footnote 4, Feynman explicitly mentioned
that he took this function from the textbook by Bode, citing its ``page
333". Remarkably, in Ref. \cite{Bode} (accounting for the 1945 edition of this textbook), page 333
is not focused on this topic. At a close inspection we can rather deduce that
Feynman borrowed formula (17-14) on page 411 of Ref. \cite{Bode}, which referred 
however \emph{not} to the gain of an amplifier, but represented
the input impedance of the compensated infinite filter composed by
$\pi$-shaped elementary blocks (notice that $\omega_0$ in the
Bode book corresponds to $2\omega_0$ in Feynman's
report).}} which he {\it assumed} to be applicable to his
theoretical reference amplifier:
\begin{eqnarray}\label{HFCresponse1}
A\left(\omega\right)=\left[\sqrt{1-\frac{\omega^2 }{4\omega_0^2 }}+\frac{i\omega }{2\omega_0 }\right]^{-k}
\end{eqnarray}
($\omega_0$ is a given parameter). For $\omega < 2\omega_0$, we have $\left| A(\omega) \right|=1$, that is a constant amplification with a  phase shift varying as $\tan^{-1}\left[ {\frac{\omega }{2\omega_0 }}/{\sqrt{1-\frac{\omega^2 }{4\omega_0^2 }}} \right]$. Conversely, for $\omega > 2\omega_0$ we have $A\left(\omega\right)=i^{-k}\left[\sqrt{\frac{\omega^2 }{4\omega_0^2 }-1}+\frac{\omega }{2\omega_0 }\right]^{-k}$, with a constant phase shift (equal to ${\pi}/{2}$). This remains true also for very large frequencies ($\omega \gg 2\omega_0$), when $A\left(\omega\right)$ behaves as $i^{-k}\left(\frac{\omega_0 }{\omega }\right)^k$. The response to the Dirac delta-function was obtained by putting the expression (\ref{HFCresponse1}) of $A\left(\omega\right)$ into Eq. (\ref{tf2}) and using contour integration techniques. The result was:
\begin{eqnarray}\label{HFResponseDelta1}
R\left(t \right)=\left( \frac{k}{t}\right) J_k\left(2 \omega_0 t \right),
\end{eqnarray}
where $J_k\left(x \right)$ is the $k^{th}$ order Bessel function.

\begin{figure}
\begin{center}
\includegraphics[height=3cm]{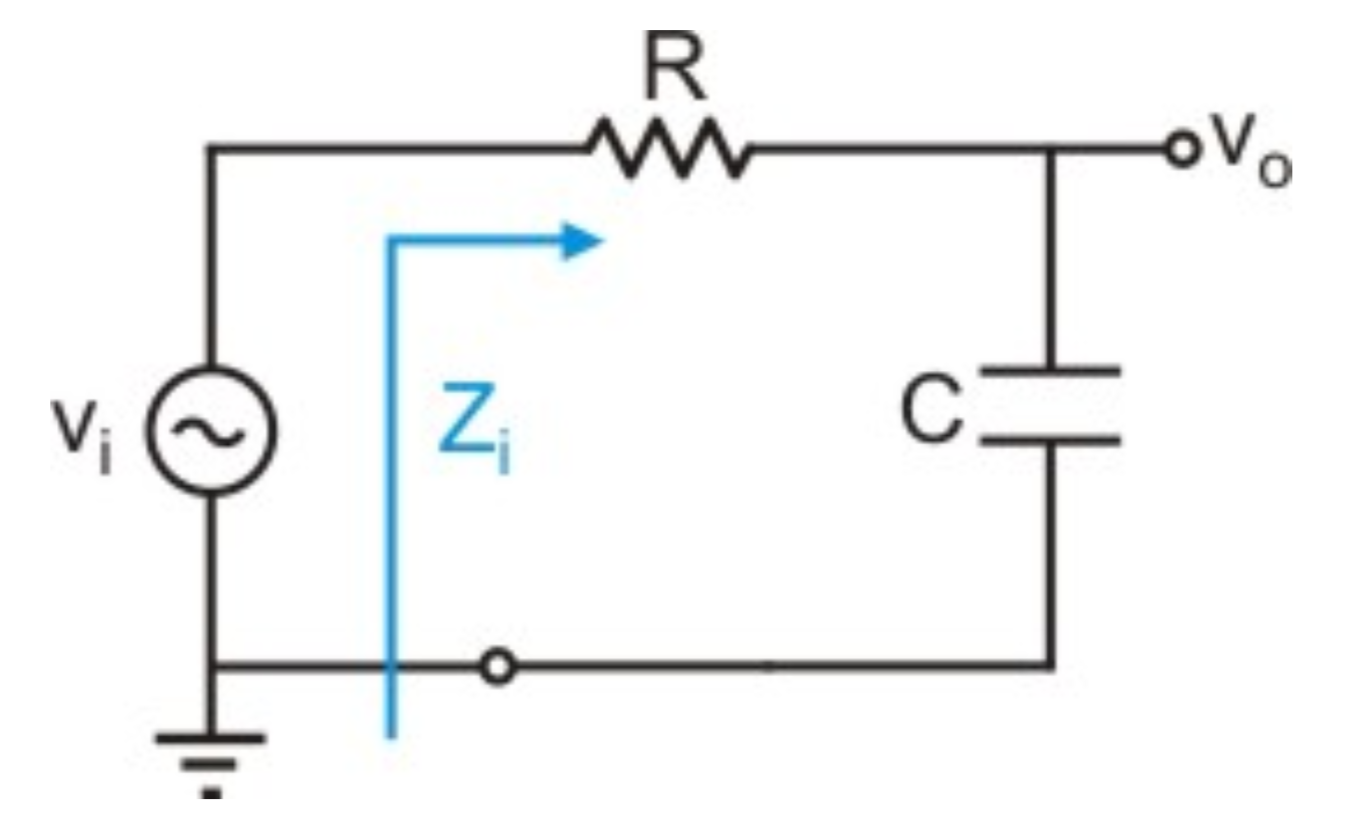}
\caption{The low-pass $RC$ ``amplifier" (filter) considered by Feynman in his report \cite{FeynmanLARep}.}
\label{fig1}
\end{center}
\end{figure}

As a second example Feynman considered another possible transfer function for his theoretical reference amplifier,
\begin{eqnarray}\label{HFCresponse2}
A\left(\omega\right)=\left[1+\frac{i\omega }{\omega_0 }\right]^{-k},
\end{eqnarray}
pointing out that, for the particular case $k=1$ and
$\omega_0={1}/{RC }$, it gives the capacitance/resistor voltage
ratio for a $RC$ circuit; and, similarly, the values
$k=2,3,4,\dots$ correspond to two, three, four, etc. series $RC$
circuits. However, the solution Feynman considered for an
amplifier holds true also for possible fractional values of $k$
(just as in the case above), although it cannot be associated to
$RC$ circuits.\footnote{It is interesting to note that similar
structures come out in the spin glass statistical theory with
the so-called Parisi replica trick \cite{Mezard}, discovered much
later than Feynman. We are grateful to Francesco Guerra for
pointing out to us this point.} Again, the response to a
delta-function pulse was worked out by putting the expression
(\ref{HFCresponse2}) of $A\left(\omega\right)$ into Eq.
(\ref{tf2}) and using contour integration techniques; the
integrand has a pole of order $k$ for $\omega=i\omega_0$, so that
by application of the residue theorem the result follows:
\begin{eqnarray}\label{HFResponseDelta2}
R\left(t \right)=\frac{\left(\omega_0 t \right)^k}{t \ \Gamma\left(k \right)} {\mathrm e}^{-\omega_0 t},
\end{eqnarray}
where $\Gamma\left(k \right)$ is the Euler gamma function.

By comparing the two examples considered,  Feynman concluded that
the amplifiers described by (\ref{HFCresponse2}) respond more slowly
than those described by (\ref{HFCresponse1}), although they are
less liable to overshoot. Many real amplifiers exhibit an
intermediate behavior, with amplification curves lying between the
above two cases.

For low-frequency cutoff amplifiers, Feynman proceeded in a similar way starting from the attenuation function
\begin{eqnarray}\label{LFCresponse1}
A\left(\omega\right)=\left[\sqrt{1-\frac{\omega_0^2 }{4\omega^2 }}-\frac{i\omega_0 }{2\omega }\right]^{-k}.
\end{eqnarray}
A constant amplification  is now obtained for $\omega > {\omega_0 }/{2 }$, while a constant phase shift for $\omega < {\omega_0 }/{2 }$. The corresponding response function for $k=1$ is, according to him:
\begin{eqnarray}\label{LFResponseDelta1}
R\left(t \right)=\delta\left(t \right)-\frac{\omega_0 }{2 }+\left( \frac{\omega_0}{2}\int_{0 }^{t }\frac{1 }{t } \, J_1 \!\! \left(\frac{\omega_0 t}{2} \right) {\mathrm d}t \right),
\end{eqnarray}
while, for $k=2$:
\begin{eqnarray}\label{LFResponseDelta1bis}
R\left(t \right)=\delta\left(t \right)-\omega_0 \, J_0 \!\! \left(\frac{\omega_0 t}{2}\right) +\left( \frac{\omega_0^2 t}{2}\right) \left[1-\int_{0 }^{t } t \,  J_1 \!\! \left(\frac{\omega_0 t}{2} \right) {\mathrm d} t \right].
\end{eqnarray}
In this case there is no high-energy cutoff, and therefore the output fully reproduces the sharpest features in input, as inferred from the presence of the delta-function in the above expressions.

\begin{figure}
\begin{center}
\begin{tabular}{cc}
\includegraphics[height=5.5cm]{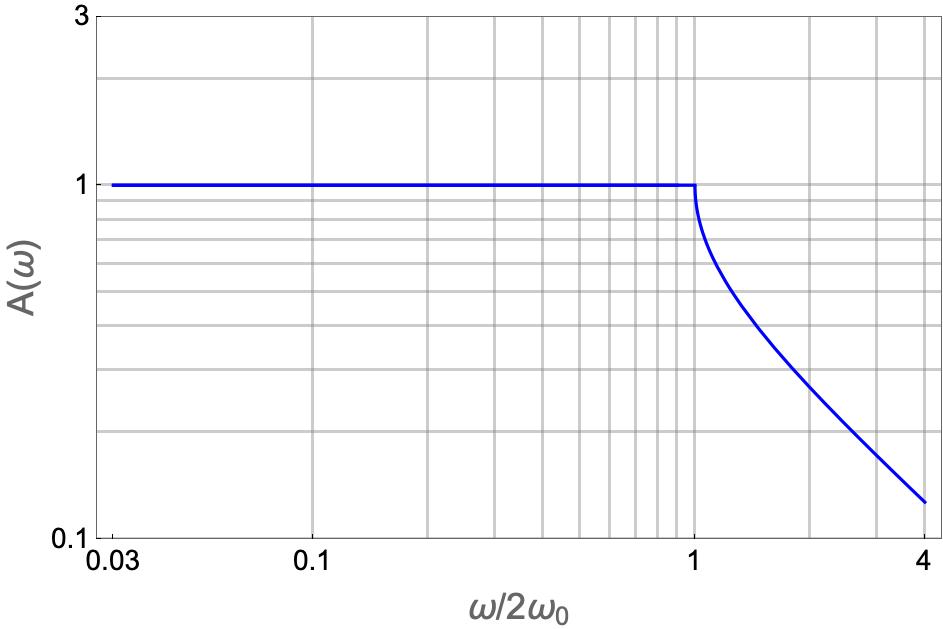}
% & b) \includegraphics[height=5.5cm]{fig2feynmampl-b.jpg}
\end{tabular}
\caption{The ``amplification" discussed by Feynman, corresponding to the magnitude of the response function in Eq. (\ref{HFCresponse1}).}
\label{fig2}
\end{center}
\end{figure}

Similarly, for the transfer function
\begin{eqnarray}\label{LFCresponse2}
A\left(\omega\right)=\left[1-\frac{i\omega_0 }{\omega }\right]^{-k},
\end{eqnarray}
characterizing, for integer values of $k$, $RC$ circuits, the corresponding response function for $k=1$ is given by:
\begin{eqnarray}\label{LFResponseDelta2}
R\left(t \right)=\delta\left(t \right)-\omega_0 \, {\mathrm e}^{-\omega_0 t},
\end{eqnarray}
while, for $k=2$, is:
\begin{eqnarray}\label{LFResponseDelta2bis}
R\left(t \right)=\delta\left(t \right)-\omega_0 \left(2- \omega_0 t \right) {\mathrm e}^{-\omega_0 t}.
\end{eqnarray}

What is the physical meaning of what Feynman obtained with such examples?

Very explicitly, in his 1946 report \cite{FeynmanLARep}, Feynman
referred to the mathematical response function $A$ as the
``amplification" of the theoretical reference amplifier he focused
on, notwithstanding the fact that, in the examples he considered,
the maximum value of such ``amplification" was just the unity.
Quite evidently, Feynman dealt  -- more properly -- with
normalized amplification, that is relative amplification with
respect to some standard, since, as noted by himself, he was ``not
interested in the absolute amplification, but only in
distortion"  {(\cite{FeynmanLARep}, on page 3)}, as
already addressed above. As a matter of fact, however, he referred
to {\it filters} (that is, passive circuital components) rather
than amplifiers, which instead involve active components.

This is particularly evident in the second kind of examples he
explicitly considered, concerning the low-pass $RC$ filter (for
$k=1$), depicted here in Figure \ref{fig1}. It is a standard
textbook exercise \cite{Valkenburg}, indeed, to show that the
expression obtained by Feynman in Eq. (\ref{HFCresponse2}) (for
$k=1$ and $\omega_0=1/RC$) just corresponds to the {\it voltage
gain} $A(\omega) = v_o/v_i$, which is then the physical meaning of
what Feynman considered the amplification in this specific
example.

A far less clear example is, instead, that corresponding to Eq.
(\ref{HFCresponse1}) (here and in the following we consider the
simplest case with $k=1$), which Feynman explicitly borrowed from
Bode \cite{Bode}, the reason being that, as we will see below, now
$A(\omega)$ {\it cannot} be interpreted as a voltage gain of some
circuit as in the other example. Nevertheless, this intricate
problem should be appropriately addressed since, as Feynman
readily realized, the ``amplification" of his ``high cut-off
frequency amplifier" is  remarkable. By plotting Eq.
(\ref{HFCresponse1}) in Figure \ref{fig2}, indeed, it is rather
evident that the theoretical reference ``amplifier" described by
it is a non-common, exceptional device, with an extremely large
pass band and a well defined falling off point. Also, as Feynman
noted, ``for very large frequencies, the amplification falls off
at the rate of 6 dB per octave (20 db per
decade)" {(\cite{FeynmanLARep}, on pages 6-7)}: for large
$\omega$, we have $|A(\omega)| \simeq \omega_0/\omega$, and thus
$|A(\omega)|_{\rm dB} \simeq 20 \log_{10} \omega_0 - 20 \log_{10}
\omega \simeq 20 \log_{10} \omega_0 - 6 \log_2 \omega$, just
confirming Feynman's statement.

In order to get some insight into the physical meaning behind Eq.
(\ref{HFCresponse1}), it is obvious to make recourse directly to
the literature quoted in the report \cite{FeynmanLARep}, that is
the book by Bode \cite{Bode}. Unfortunately, although Bode was the
well-known engineer who pioneered the modern control theory and
electronic communications, he expressed himself rather as a
mathematician in the book mentioned, where the correspondence
between formulas and real world was mostly disregarded or even
missing. It is a matter of fact, indeed, that parts of his book
are cryptic or nebulous, just including the pages dealing with the
high cut-off frequency interstage circuits. Fortunately enough,
conversely, the original work to which Bode refers, i.e. the
important paper by H.A. Wheeler \cite{Wheeler}, does not suffer
from this, so that some hints can be nevertheless usefully gained.

\begin{figure}
\begin{center}
\begin{tabular}{c}
a) \includegraphics[height=3cm]{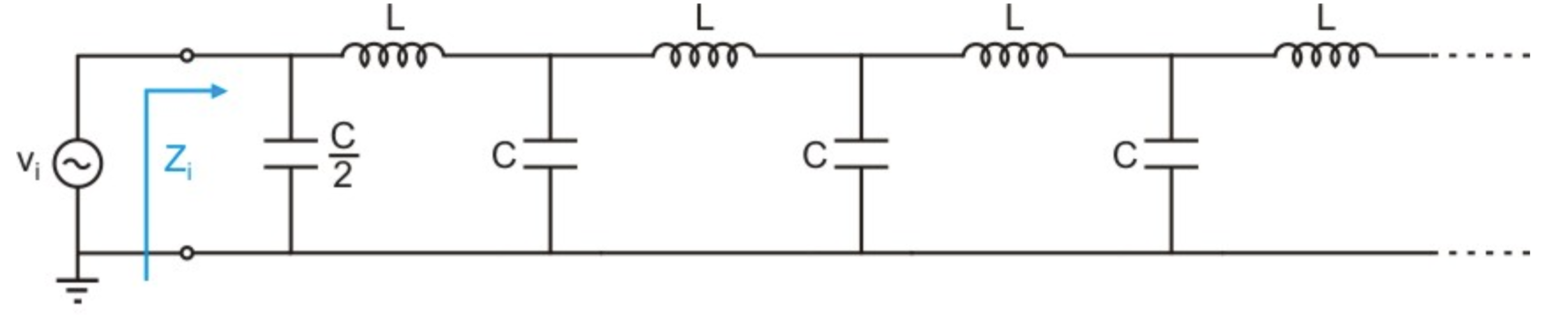}
\\ \\
b) \includegraphics[height=3.5cm]{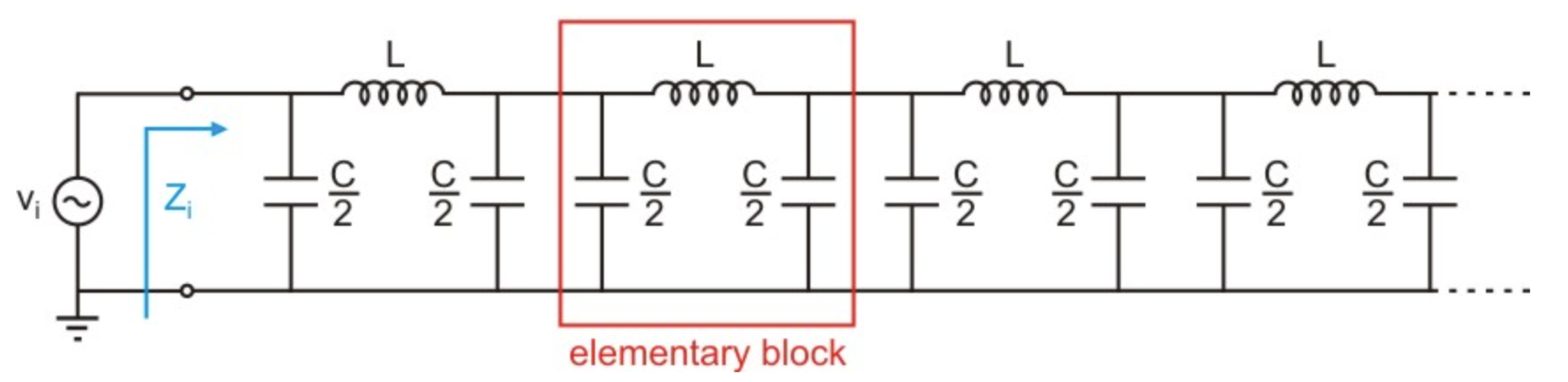}
\\ \\
c) \includegraphics[height=3.2cm]{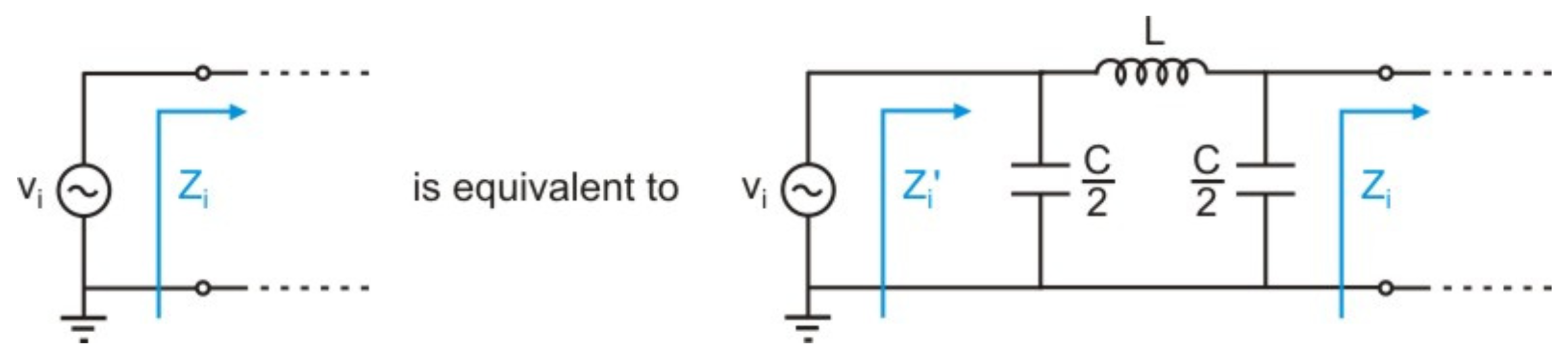}
\\ \\
d) \includegraphics[height=3cm]{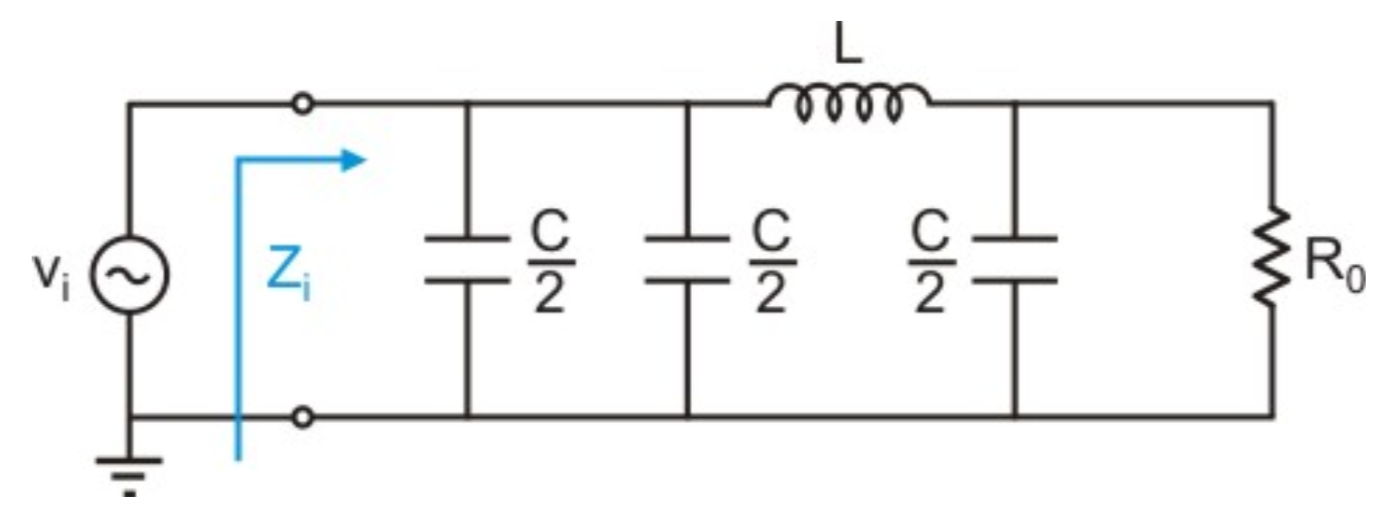}
\end{tabular}
\caption{Infinite low-pass $LC$ filter: a) basic representation; b) infinite replica of a $\pi$-shaped elementary block; c) circuit evaluation strategy; d) approximation of the (compensated) infinite filter with a finite network with one elementary block (see text).}
\label{fig3}
\end{center}
\end{figure}

\begin{figure}
\begin{center}
\begin{tabular}{ccc}
a) \includegraphics[height=3cm]{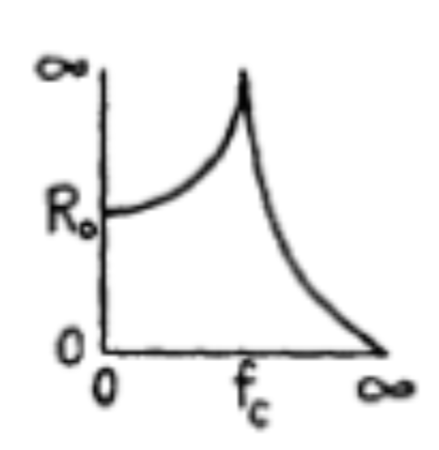}
& $\qquad \qquad$ &
b) \includegraphics[height=3cm]{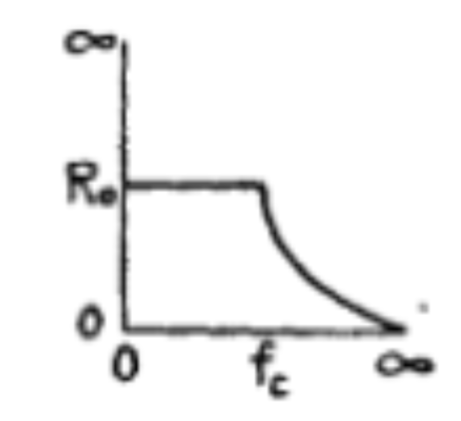}
\end{tabular}
\caption{Magnitude of the input impedance of the low-pass filter
{as a function of frequency ($f_c={\omega_c}/{2\pi}$)}: a)
presence of a resonance, as from Eq. (\ref{impLC}); uniform
behavior, for the filter in Figure \ref{fig5}b (original figures
taken from Ref. \cite{Wheeler}, {page 430}).}
\label{fig4}
\end{center}
\end{figure}

\begin{figure}
\begin{center}
\begin{tabular}{c}
a) \includegraphics[height=3cm]{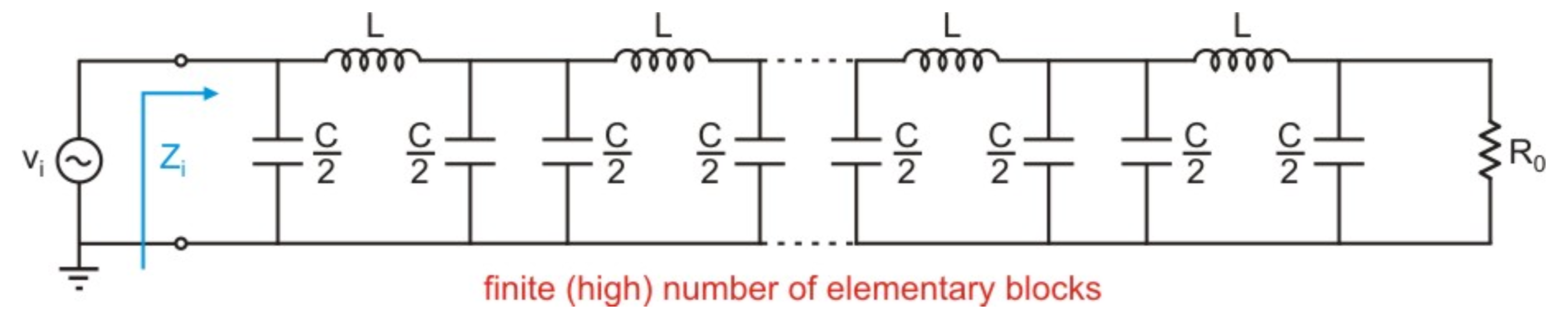}
\\ \\
b) \includegraphics[height=2.5cm]{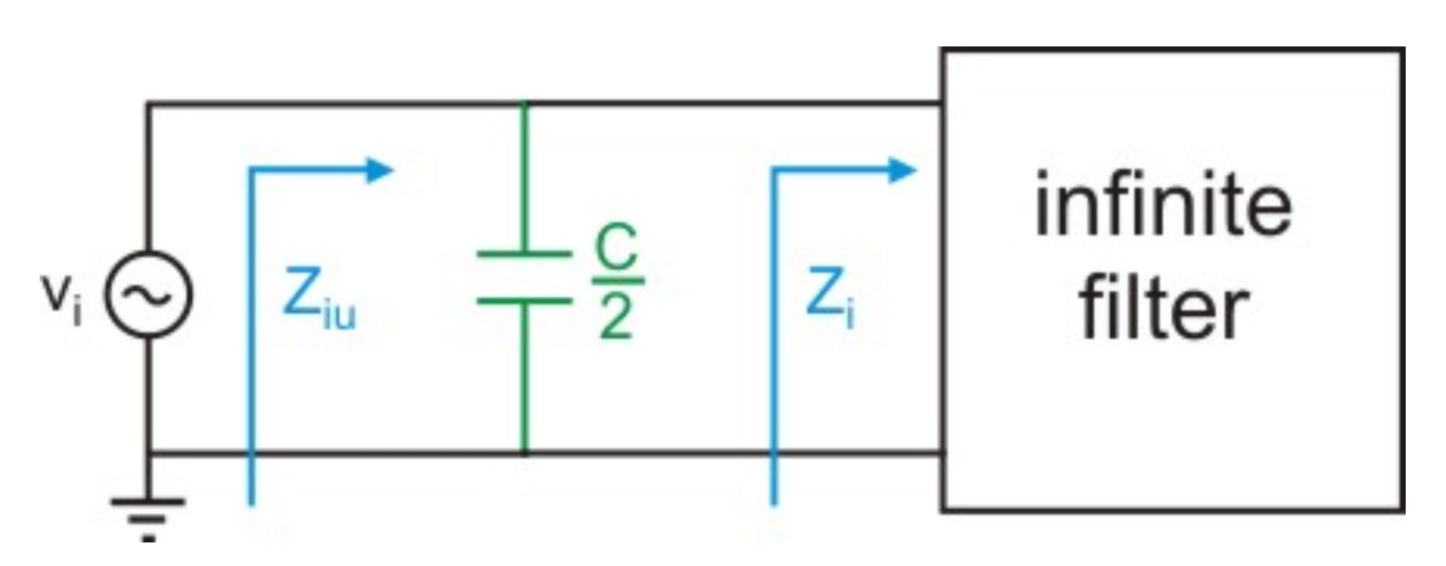}
\end{tabular}
\caption{Approximations of an infinite low-pass $LC$ filter: a) finite network with a terminating characteristic resistance; b) compensated infinite filter with a shunt capacitance. }
\label{fig5}
\end{center}
\end{figure}

Let us start with a (mathematical) exercise worked out by Feynman
in his most famous {\it Lectures on Physics} \cite{FeynmanBook},
that is the infinite low-pass $LC$ filter in Figure \ref{fig3}a.
Such a filter can be represented as shown in Figure \ref{fig3}b,
and can be viewed as the infinite replica of the $\pi$-shaped
elementary block shown. The {\it input impedance} $Z_i$ of the
circuit can be determined by invoking a renormalization-like
strategy: since the filter is infinite, $Z_i$ is not modified by
adding a further elementary block before $Z_i$, as shown in Figure
\ref{fig3}c, that is $Z^\prime_i = Z_i$. From this equation, after
some algebra the {\it characteristic impedance} (the impedance
viewed before any block of the infinite circuit) can be deduced
for the filter above, known in the literature as constant-$k$
filter (where $k=\sqrt{L/C}$):
\begin{eqnarray}\label{impLC}
Z_i(\omega) = \frac{R_0}{\displaystyle \sqrt{1 - \frac{\omega^2}{\omega_c^2}}}
\end{eqnarray}
($R_0= \sqrt{L/C}$ and $\omega_c =2/\sqrt{LC}$, using the
nomenclature in \cite{Wheeler}). $Z_i(\omega)$ is imaginary
(purely reactive) for $\omega > \omega_c$, and its magnitude is
depicted in Figure \ref{fig4}a (as also shown in \cite{Wheeler}).
Bode \cite{Bode} referred to such an infinite filter (which was,
nevertheless, not made explicit in the text) as a cryptic ``ideal
low-pass structure", but an intriguing statement was made: ``for
practical purposes the ideal low-pass structure can be
approximated by a finite network giving a reasonably accurate
match to the terminating resistance equal to the input
(characteristic) impedance at $\omega=0$, $R_0=\sqrt{L/C}=
2/\omega_c C$'' {(\cite{Bode}, on page 405)}. That is,
the infinite filter in Figure \ref{fig3}c can be approximated by
the finite one in Figure \ref{fig5}a, as we indeed checked (for a
large number of elementary blocks) by resorting to the popular
circuit simulation program PSPICE \cite{pspice}.

\begin{figure}
\begin{center}
\includegraphics[height=6cm]{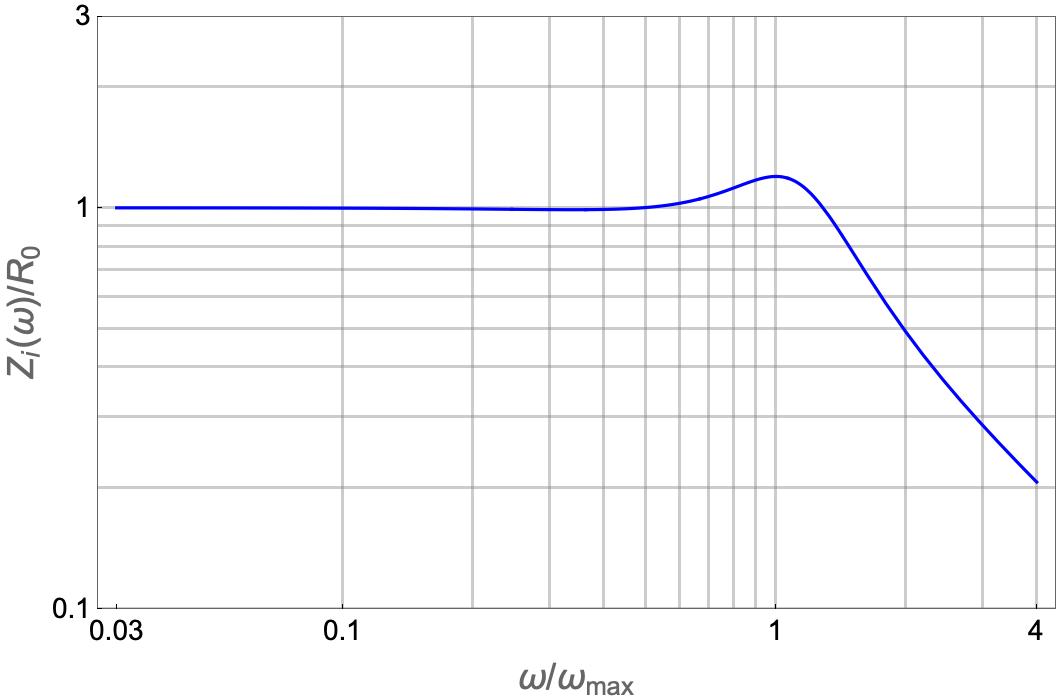}
\caption{Normalized magnitude of the input impedance of the filter in Figure \ref{fig3}d as a function of the frequency (normalized to the frequency of the peak).}
\label{fig6}
\end{center}
\end{figure}

The presence of a resonance for a given frequency is, of course,
an evident problem for a filter -- though expected for an {\it
infinite} filter --, as pointed out by Wheeler, who then noticed
that ``the input impedance can be made uniform over the pass band
by adding another shunt capacitance $C/2$ {\it before} the
infinite filter" {(\cite{Wheeler}, on page 430)}, as illustrated
here in Figure \ref{fig5}b. Such impedance can be quite easily
calculated \cite{Bode, Wheeler}, the result being the following:
\begin{eqnarray}\label{inputimp}
Z_i(\omega) = \frac{R_0}{\displaystyle \sqrt{1-\frac{\omega^2 }{\omega_c^2 }}+\frac{i\omega }{\omega_c }} \, .
\end{eqnarray}
Its magnitude $|Z_i(\omega)|$ for $\omega < \omega_c$ has indeed a
uniform behavior over the pass pand, while for $\omega > \omega_c$
decreases for increasing frequency, as shown in Figure \ref{fig4}b
\cite{Wheeler}. A general result  of such a kind was reported also
by Bode, but here the {above cited} cryptic statement that
``for practical purposes the ideal low-pass structure can be
approximated by a finite network giving a reasonably accurate
match to the terminating resistance" \cite{Bode} seems to allude
that suitable methods exist that allow constructing also very
compact (and special) networks  exhibiting the above $Z_i$s. And,
in fact, examples are reported in his textbook, the first of them
promoting the adoption of only one elementary block as in Figure
\ref{fig3}d. The evaluation of this circuit effectively leads to a
situation similar to that envisaged by Wheeler, as can be seen in
Figure \ref{fig6}, although the uniform low-pass behavior is now
not so well marked as above.

The similarity between Feynman's Eq. (\ref{HFCresponse1}) and
Wheeler-Bode's Eq. (\ref{inputimp}) then allows to speculate again
about the physical interpretation of Feynman's ``amplification"
$A(\omega)$ in terms of a voltage gain, just as in the other ($RC$
circuit) example mentioned above. However, it is quite important
to note that the behavior of the magnitudes of the voltage gains
of the compensated version of the finite filter in Figure
\ref{fig5}a and its single-block variant in Figure \ref{fig3}d are
{\it different} from the magnitude of $A(\omega)$ in Eq.
(\ref{HFCresponse1}). Also, in general the behavior with the
frequency of the voltage gain is different from that of an input
impedance. Unfortunately, then, this definitively leads to the
conclusion that Feynman's  ``amplification" $A(\omega)$ in Eq.
(\ref{HFCresponse1}) does {\it not} represent the voltage gain of
finite filters approximating the infinite compensated filter in
Figure \ref{fig5}b, notwithstanding Feynman's explicit reference
to Bode's (cryptic) result.

{There are two possible interpretations of this mismatch.
Feynman may have misinterpreted Eq. (\ref{HFCresponse1}), thinking
that it described the voltage gain of a real device, or may have
been aware of this fact, but nevertheless studied it since it
met the causality constraints.}

\section{Conclusions}

In the present paper we have analyzed Feynman's work on amplifier
response performed at Los Alamos and described  in a technical
report of 1946, as well as lectured on at the Cornell University
in 1946-47 during his course on Mathematical Methods. Quite
interestingly, such a work later flowed in the Hughes lectures on
Mathematical Methods in Physics and Engineering of 1970-71, where
Feynman also remarked on causality properties. Inspiration for
such a work was given by his involvement in the Manhattan Project
particularly connected with the experiments performed, where the
necessity emerged of feeding the output pulses of counters into
amplifiers or several other circuits, with the risk of
introducing distortion at each step. In order to address such an
issue, Feynman conceived a theoretical ``reference amplifier" able
to provide a useful standard in practical comparison with real
devices. A general theory was then elaborated, as described here
in Sect. 2, having at its core the response function $R(t)$ of
that amplifier (assumed to be linear), from which he was able to
deduce the basic features of an amplifier just from its response
to a pulse or to a sine wave of definite frequency. The main
properties of the response function were explicitly worked out,
and a particular reference was given to the causality issue, that
is to certain relations between frequency and phase shift that a
real amplifier has to satisfy in order not to allow output signals
to appear before input ones. As shortly pointed out in Sect, 3, to
this regard Feynman interestingly introduced dispersion relations
to be satisfied by the response function, probably inspired by
similar issues in different branch of Physics (and probably
inspiring his later contributions in Quantum Electrodynamics).

Feynman's analysis, however, was not limited to a general theory
but, in order to apply it to practical problems, a couple of
remarkable examples were  in addition worked out, as discussed in
Sect. 4, both for high-frequency cutoff amplifiers and for
low-frequency ones. Quite interestingly, the reference textbook
for his study was the famous mathematically-oriented book by Bode
\cite{Bode}, which certainly did not provide practical, physical
insight into the problem. Likely, this was the basis for the
misunderstanding that is apparent when inspecting in detail
Feynman's report, where, irrespective of the fact that he
explicitly dealt with amplifiers, appropriate reference should be
to filters, that is to passive -- rather than active -- circuital
components. Indeed, one of the two different examples Feynman
considered referred to the standard low-pass $RC$ filter, from
which study we have been able to deduce the physical meaning of
the mathematical response function associated to the (relative)
``amplification" of the ``amplifier", which can then be
interpreted as the voltage gain of the filter.

However, the particularly mysterious, yet interesting,  second
example considered by Feynman (again borrowed from Bode) certainly
deserves a special mention, if only for the exceptional features
of the device possibly described by Eq. (\ref{HFCresponse1}), with
an extremely large pass band and a well defined falling off point.
Indeed, the mystery comes from the fact that, as carefully
explained in the previous section, it seems that no real device
can possess the exceptional features referred above, the
``amplification" considered by Feynman not being able to be
associated to the voltage gain of real filters. It is then not
easy to understand the reasons behind the choice of Eq.
(\ref{HFCresponse1}), which, however, {in Feynman's
intentions, should represent} a voltage gain {instead}.

Of key relevance in this case is certainly Feynman's reference to
Bode's textbook,  within which the correspondence between formulas
and real world is mostly disregarded or even missing. Being not an
expert in the field, Feynman may well have misinterpreted Eq.
(\ref{HFCresponse1}) as the voltage gain of the compensated
version of the finite filter in Figure \ref{fig5}a or of its
compact variants, like the single-block filter in Figure
\ref{fig3}d, notwithstanding the fact that Eq.
(\ref{HFCresponse1}) gives the impedance $Z_i$ of the compensated
infinite filter normalized to its pass-band value $R_0$. On the
other hand, it is also possible that Feynman was aware that Eq.
(\ref{HFCresponse1}) did not correspond to any real
filter/amplifier -- although he stated that Eq.
(\ref{HFCresponse1}) ``satisfies conditions for the existence of a
corresponding real amplifier" {(\cite{FeynmanLARep}, on page 6)}
-- and presented it
as the normalized {\it loop gain} ({\it return ratio}) of an {\it
ideal} feedback amplifier (the loop gain is, indeed, the product
between the gain of the feedback-less amplifier, i.e. open-loop
gain, and the feedback factor, i.e. gain of the feedback block).
{One could speculate that} such a possibility comes from
the consideration that feedback amplifiers were deeply studied by
Bode and Nyquist at the Bell labs during the 1930s.\footnote{We
are very grateful to Prof. Codecasa for pointing out to us this
point, and the corresponding possible interpretation.} Of course,
Feynman assumed his feedback amplifier with the best stability
behavior, i.e. large pass band and {\it constant} phase (equal to
$\pi/2$ in the present case) in the attenuation band, and he may
therefore have hoped that something similar could have been
obtained with a compact filter synthesized with special
techniques. However, we cannot but observe that this is not
evident in the pages of Bode's book cited by Feynman
\cite{FeynmanLARep} and, though it is possible that he did not
even care about the existence of the real circuit, it is a matter
of fact that the feedback amplifier is ideal, and does not
correspond to any real amplifier.

The mystery on this point, then, still remains, but it is quite
remarkable that -- once more -- the intriguing personality emerges
of a multifaceted scientist, who {unleashed} his
unconventional genius even in unexpected fields of application.

\end{document}